\documentclass[conference]{IEEEtran}
\IEEEoverridecommandlockouts
\usepackage{cite}
\usepackage{amsmath,amssymb,amsfonts}
\usepackage{algorithmic}
\usepackage{graphicx}
\usepackage{textcomp}
\usepackage{xcolor}
\def\BibTeX{{\rm B\kern-.05em{\sc i\kern-.025em b}\kern-.08em
    T\kern-.1667em\lower.7ex\hbox{E}\kern-.125emX}}

\begin{document}

\title{A Self-Paced Generative AI Chatbot Approach for Programming Skills Training}

\author{\IEEEauthorblockN{M.A.F. Aamina\IEEEauthorrefmark{1}, 
        V. Kavishcan\IEEEauthorrefmark{1},
        W.M.P.B.B. Jayaratne\IEEEauthorrefmark{1},
        K.K.D.S.N. Kannangara \IEEEauthorrefmark{1},
        A.A. Aamil \IEEEauthorrefmark{1},
        Achini Adikari\IEEEauthorrefmark{2}
			}
    \IEEEauthorblockA{\IEEEauthorrefmark{1}School of Computing, Informatics Institute of Technology, Sri Lanka\\
    \IEEEauthorrefmark{2}Centre for Data Analytics and Cognition, La Trobe University, Australia\\ 
    }}

\maketitle

\begin{abstract}
Computer programming represents a rapidly evolving and sought-after career path in the 21st century. Nevertheless, novice learners may find the process intimidating for several reasons, such as limited and highly competitive career opportunities, peer and parental pressure for academic success, and course difficulties. These factors frequently contribute to anxiety and eventual dropout as a result of fear. Furthermore, research has demonstrated that beginners are significantly deterred by the fear of failure, which results in programming anxiety and and a sense of being overwhelmed by intricate topics, ultimately leading to dropping out. This project undertakes an exploration beyond the scope of conventional code learning platforms by identifying and utilising effective and personalised strategies of learning. The proposed solution incorporates features such as AI-generated challenging questions, mindfulness quotes, and tips to motivate users, along with an AI chatbot that functions as a motivational aid. In addition, the suggested solution integrates personalized roadmaps and gamification elements to maintain user involvement. The project aims to systematically monitor the progress of novice programmers and enhance their knowledge of coding with a personalised, revised curriculum to help mitigate the fear of coding and boost confidence. \\
\end{abstract}


\section{Introduction}
Computer programming often known as coding, is an essential skill for future generations in the 21st century \cite{b1}. Over the past two decades, the internet and mobile technologies have catalyzed remarkable advancements in various domains of computer science. The responsibilities of computer science, such as automating processes, facilitating communication, providing better products and services, and contributing to global productivity, have led to an increased dependence on computer software by human beings \cite{b2}. Due to these adaptations, there is a demand for individuals proficient in software development globally. While there is an argument that teaching student’s computer coding encourages them to solve problems in general and could aid them in the development of cognitive abilities beyond coding, many computer science students find programming to be challenging and intimidating \cite{b3}. 

There has been a significant amount of research to determine the factors that contribute to students’ success in programming. The findings indicate that a significant number of students often quit or struggle to achieve success in programming. Experiencing anxiety about programming emerges as a substantial element that can predict an individual’s performance in programming tasks. Previous studies suggests that factors such as familiarity with programming, mathematical knowledge, problem solving skills, learning styles, student expectations for outcomes, comfort levels and self-efficacy can influence how well students perform in programming \cite{b4}. 

According to Connolly, Murphy, and Moore \cite{b5}, programming anxiety is a psychological state that arises in specific situations due to bad experiences or expectations related to computer programming. Scott \cite{b6} states that students frequently experience programming anxiety during the early phases of programming courses due to the introduction of completely new concepts and materials. In addition, the level of abstraction in introductory programming courses has risen in recent decades, leading to an increase in programming anxiety \cite{b5}. Chang \cite{b7} used a sample of 307 people to examine the connection between the perceived difficulty of programming tasks and programming anxiety. The data revealed a significant correlation between these two variables, suggesting that as the perceived difficulty of programming tasks grew, students' levels of programming anxiety also increased.

Conventional coding education platforms often rely on theoretical methods lacking in engagement, resulting in a loss of interest and dropout from the learning process \cite{b8}. Jiang, Zhao, Wang, and Hu \cite{b9} reveal that dropout rates on code learning platforms have reached up to 90\%. Students who discontinue online courses may become frustrated and lose confidence, discouraging them from attempting further enrollment. If students fail to complete their initial online courses, they may face difficulties in enrolling in additional distance courses. Additionally, this can negatively impact the student’s self-confidence \cite{b10}. 

Khan Academy is an online learning platform that offers self-paced courses based on human expertise and motivation. The platform covers a wide range of subjects, including mathematics, science, economics, humanities, and computer programming \cite{b11}. Users initiate questions, receive responses, and can provide feedback by up-voting or down-voting content. Analysis of user interactions provides insight into activity patterns, particularly between experts and novice learners. However, as the curriculum becomes more demanding, most low-performing users drop out, while high-performing experts persist. The smaller community tends to engage more actively than before, with some users dropping out due to lack of expertise. Mondal etal \cite{b12} suggests that users who drop out may be novice learners who ask multiple questions but lack the required expertise to answer many questions themselves \cite{b12}.

Codecademy is a popular online MOOC that offers free, interactive lessons for various programming topics \cite{13}. Shen and Lee \cite{b14} have raised complaints about Codecademy. They found that the codes extracted were not practical and too rigid, leading some learners to be confused after finishing courses. This confusion was primarily due to the lack of practical projects during the learning process and the content focusing on aspects such as syntax rather than problem-solving skills. Codecademy offers structured tutorials where learners follow a strict, predefined path, which some learners find helpful, while others find it repetitive and boring \cite{b14}. 
\vspace{0.5\baselineskip} 

Thus, the proposed solution aims to directly address these concerns. A specialized and individualised code learning platform is being developed to transform the learning experience into an enjoyable and successful journey. It will present essential concepts in a simplified manner to help build confidence, with mindfulness approaches to motivate users and encourage them to persist in their learning journey. Designed specifically for beginner programmers, this forum serves as a resource for refreshing their knowledge and overcoming challenges when encountering difficulties with a specific topic. The primary goal is to support beginner coders who may be tempted to give up due to coding-related fears arising from various reasons.

\section{Methodology}
\subsection{Key functionalities and Implementation}
The architecture of Accodemy, illustrated in Figure 1, adopts a monolithic three-tier design. This structure encompasses the data layer, the Generative AI layer, and the output layer. The data layer hosts datasets and other pertinent data storage resources. Data is then processed in the Generative AI layer, where it is transformed into dynamic questions, personalized roadmaps, and motivational tips.

\begin{figure}[htbp]
\centerline{\includegraphics[width= 6 cm, height= 10 cm]{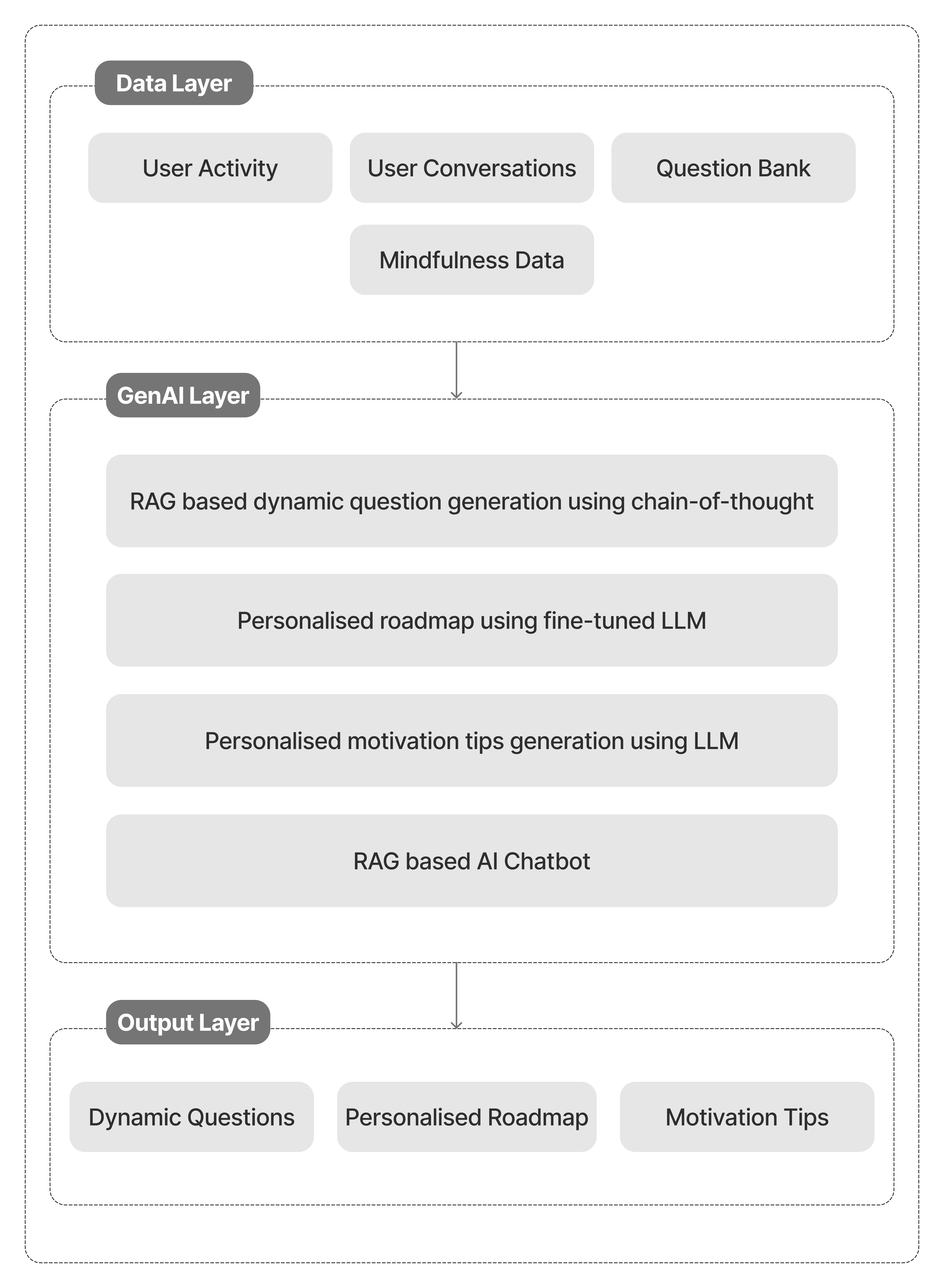}}
\caption{High level architecture diagram of the proposed system}
\label{figure}
\end{figure}

The prototype is being developed using the MERN stack. The user engages with the system through a web interface that has been created with React.js in the front end. The data storage and management are handled by MongoDB, while the backend is built using Express.js and Node.js. Clerk is utilized for authentication, while Large Language Models (LLM) have been used to personalize the learning experience for each user. 

\subsection{Use of Generative AI}
Generative AI is a form of artificial intelligence that uses trained data models to create new content like text, images, audio, and videos. It differs from traditional methods that process or analyze existing data by learning patterns and training on extensive datasets. Generative AI is used in fields like art, gaming, and entertainment, and is expanding into healthcare, manufacturing, and finance. Its application is based on learning patterns and generating outputs resembling the training data, making it a novel approach to generate fresh and resourceful content \cite{b15}.

\paragraph{Rag Model and LangChain Implementation}
The Retrieval-Augmented Generation (RAG) model is a text generation tool that uses source data to generate desired text. The process involves collecting and preparing relevant source data, breaking it into smaller chunks, embedding them into meaningful vector representations, and building a vector database. The model then searches for relevant information within the database, decoding the retrieved chunks back into original text data, and generating the text. Users can specify the type, length, and linguistic style of the text. The RAG model aims to produce more accurate and meaningful text outputs \cite{b15}.

\begin{figure}[htbp]
\centerline{\includegraphics[width=9cm, height=6.5cm]{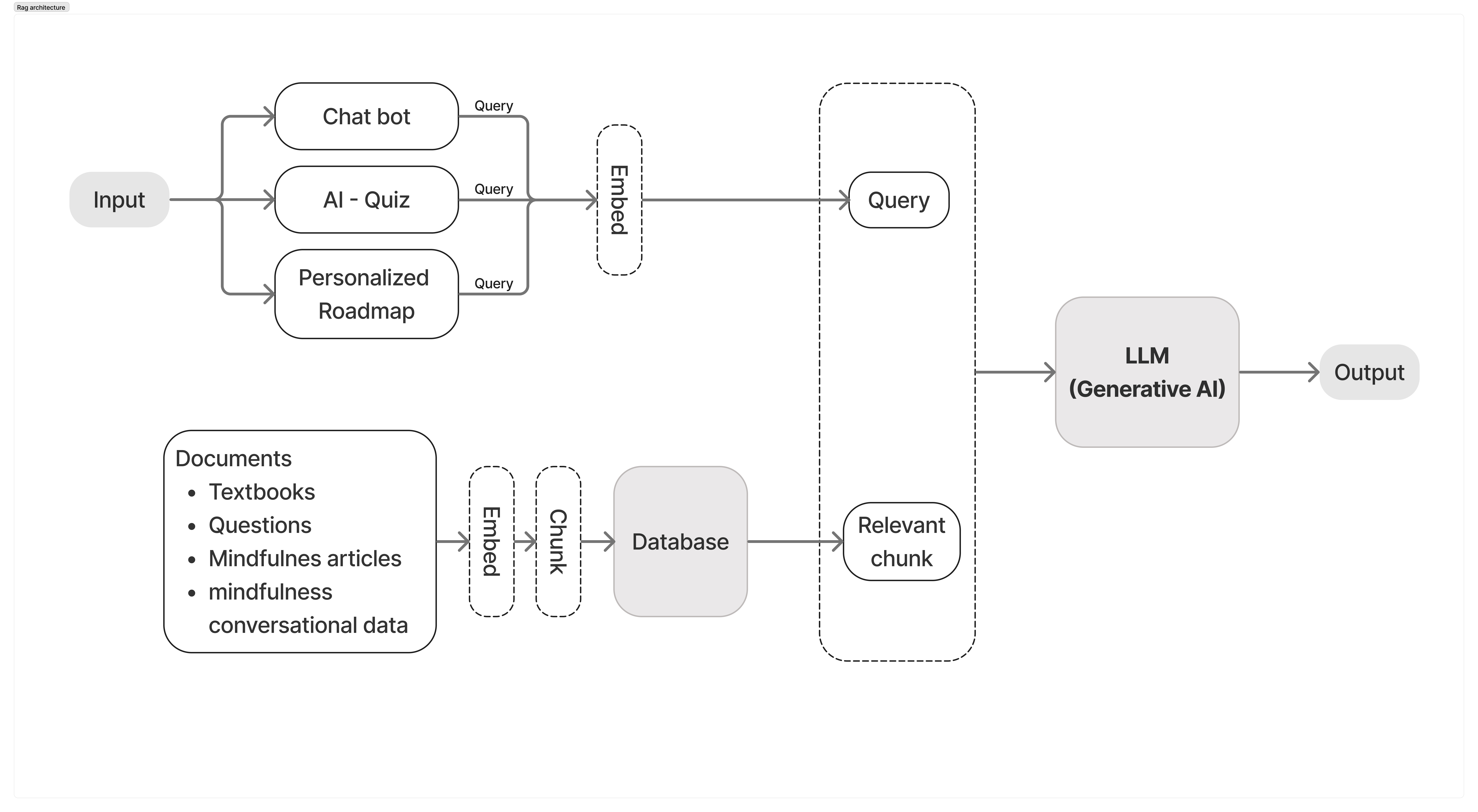}}
\caption{Rag model architecture}
\label{figure}
\end{figure}

LangChain is an open-source framework that uses LLM to perform various natural language processing tasks, including translation, summarization, question answering, text generation, and natural language inference \cite{b15}.
The implemented prototype utilises the Generative AI functionalities of LLMs derived from both proprietary models produced by OpenAI and open-source models. During the development process, the LangChain framework is employed, while the GPT 3.5-turbo model is used to perform particular tasks like chunking and embedding. 

MongoDB is used as the vector repository to store high-dimensional data, such as embedding and vectors. The selection of MongoDB was based on its straightforward deployment and effective management of intricate data structures, rendering it well-suited for the storage of inquiries, knowledge bases, and datasets within the prototype.

\begin{figure}[htbp]
\centerline{\includegraphics[width=7.5cm, height=2cm]{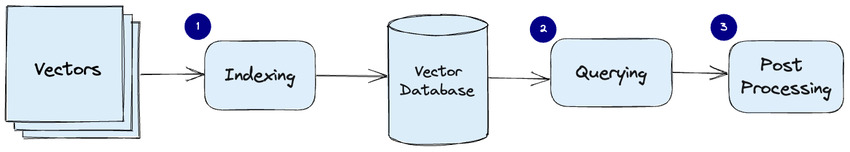}}
\caption{Pipeline for a vector database \cite{b15}}
\label{figure}
\end{figure}

\textbf{Dynamic Questions}: The dataset collected using web scraping techniques, which includes questions with explanations and answers, is embedded in the MongoDB vector database. In the project, The Chain of Thought (CoT) is employed, providing examples of how questions should be generated within the prompt, which are then used to guide the LLM in generating similar questions. The CoT method guides LLMs through prompts, enabling them to construct answers step by step by connecting information, resulting in precise responses for complex questions\cite{b15}. When generating dynamic questions, the RAG model searches for cosine similarity between the dataset and the personalized query, retrieving relevant questions. These questions, along with relevant prompts, are then forwarded to the LLM. 

The cosine-similarity refers to the cosine of the angle formed by two vectors, or alternatively, the dot product obtained by multiplying their normalizations \cite{b16}. It is a statistical measure employed to quantify the extent of similarity between two documents, with a specific focus on phrases. It aims to find similarities between vector-vector query documents, allowing for a gradual comparison of the two. The method involves scalar multiplication between the query and the document, squared document length, and root of rank two, followed by long multiplication with the outcome document and the query \cite{b17}. Furthermore, cosine similarity search has been used to effectively identify instances of plagiarism within the extensive dataset\cite{b18}.

\begin{equation}
\cos \theta = \frac{\mathbf{A} \cdot \mathbf{B}}{\| \mathbf{A} \| \times \| \mathbf{B} \|} = \frac{\sum_{i=1}^{n} a_i b_i}{\sqrt{\sum_{i=1}^{n} a_i^2} \times \sqrt{\sum_{i=1}^{n} b_i^2}}
\end{equation}

Let $\theta$ represent the angle formed by vectors A and B in an n-dimensional space. The dot product of A and B is denoted as A·B. The lengths of A and B are represented by $\|A\|$ and $\|B\|$, respectively. The i-th element of A and B is denoted as $a_i$ and $b_i$. Assuming that both $a_i$ and $b_i$ are positive, the COS scores are constrained within the range of 0 to 1. A COS value of 0 indicates that the two texts are entirely dissimilar, indicating that the two vectors are orthogonal. Contrarywise, a COS value of 1 signifies that the two vectors are identical in terms of word composition, indicating that they overlap but may have varying lengths \cite{b19}.

This method enables RAG to pre-inform the LLM about relevant queries and question datasets, thereby reducing hallucination tendencies and enhancing response accuracy. These dynamic questions are intentionally designed to be more challenging than the static questions initially attempted by the user, aiding in further refining their knowledge.
In order to create a sample question dataset that aligns with the objectives of this study, the author employed web scraping techniques to collect pertinent data. The dataset functions as the fundamental basis for training the knowledge base, which dynamically generates questions specific to the project’s requirements. 

This feature also includes question explanations generated by the LLM. When users encounter difficulty while attempting questions, they can access these explanations for clarity and guidance, aiding in understanding the concepts and solving the questions more effectively. Subsequently, upon completing the quiz or questions, users will receive motivational quotes generated by the LLM, congratulating them for good results or offering encouragement in case of failure, based on their performance. These quotes aim to encourage and inspire users further in their learning journey. This feature creates mindfulness to motivate novice learners to continue their learning within the proposed platform.
\vspace{0.5\baselineskip} 

\textbf{AI Chatbot}: The chatbot utilizes an  RAG model, integrating a provided knowledge source within its framework. By accessing both the knowledge source and the OpenAI LLM, it delivers highly accurate responses with minimal token usage. Its primary function is to serve as a supportive assistant, providing aid to individuals grappling with coding-related anxiety and promoting mental well-being. Through compassionate interaction, the chatbot offers guidance and practical mindfulness techniques to help users manage their anxiety and enhance their overall mental health.

Currently, the training of the AI chatbot involves using the "Chatbot for Mental Health Conversations" dataset, which is available on Kaggle under the title "Mental Health Conversational Dataset." This dataset, curated by Jocelyn Dumlao, provides a valuable collection of informative and pre-processed data specifically designed for mental health conversations \cite{b20}. However, in subsequent development stages, a dataset constructed specifically to coding anxiety will be utilized to further refine the RAG model. This approach aims to make the AI chatbot more attuned and precise in its responses, enhancing its effectiveness in providing mindful support to users. 
\vspace{0.5\baselineskip} 

The prototype additionally incorporates supplementary functionalities that utilize generative AI to enhance the user experience and establish an inclusive learning environment.
\vspace{0.5\baselineskip} 

\textbf{Personalized Roadmap}: The user has the capability to create a roadmap by selecting specific parameters such as timeline, topics, and programming language preference. Using this selected information, the fine tuned LLM generates a user-friendly and practical roadmap tailored to the user's preferences. This roadmap includes chosen lessons and timelines, providing a structured guide for the user's learning journey. This feature offers a distinctive customization of the users' learning path based on their individual preferences.
\vspace{0.5\baselineskip} 

\textbf{Motivational Tips}: The ``Tip of the Day' feature utilizes the LLM to generate quotes and tips based on the user’s performance and interactions. When a user signs up and logs in to the system, their progress, quiz results, and the time taken to complete quizzes will be tracked and saved in the database. By analyzing the user’s behavior, progress, and preferences, the LLM provides personalized daily insights aimed at uplifting and motivating the user. These quotes and tips serve as gentle reminders or encouragements, helping the user stay focused, inspired, and engaged in their daily activities or goals. This feature stands out as a mindfulness approach in the proposed solution. 

\section{Implementation Results}
The solution and the development framework applied follow the high-level architectural diagram of the system illustrated in Figure 1. The minimum viable product (MVP) of project Accodemy has been implemented, incorporating Generative AI. Through these implementation results, the methods of implementation and the factors needed to be considered during the development process of the features incorporating generative AI are discussed as follows.

The generation of dynamic questions plays a critical role in the implementation process, as it carries significance in the incorporation of generative AI, hence enhancing novelty. Figure 4 demonstrates the questions created by the LLM. The aforementioned AI questions dataset includes both the difficulty level and the type of lesson topic. The data is kept within the Vector repository, operating on the MongoDB platform. MongoDB's vector search functionality uses K-nearest neighbor classification and cosine similarity to retrieve the data. The retrieval process is significantly influenced by the quality of chunking, specifically in relation to the embeddings of the chunk. This, in turn, impacts the similarity and matching of chunks to user queries. The questions are created based on the specified parameters and prompt, considering the lesson topic and difficulty level.

\begin{figure}[htbp]
\centerline{\includegraphics[width= 9cm, height=3cm]{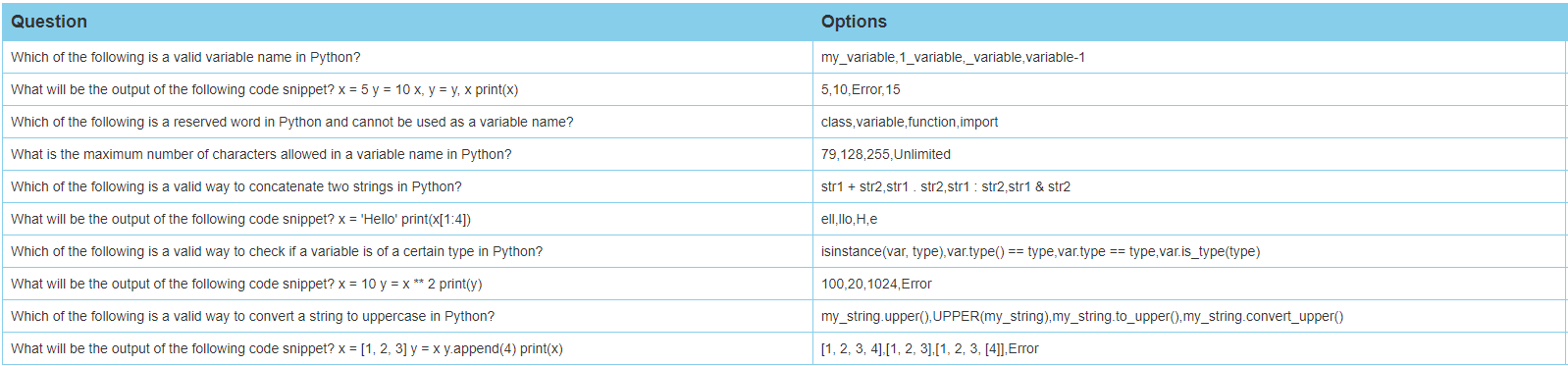}}
\caption{LLM generated dynamic questions}
\label{figure}
\end{figure}

An advantage of incorporating a RAG-based model into the LLM is the ability to synthesize contextually relevant, accurate, and up-to-date information by merging retrieval methods with the generative capabilities of LLMs. The Rag system integrates the information retrieval capabilities and generative prowess of LLMs. The retrieval component is responsible for extracting relevant information from the data store in response to a user query \cite{b21}. 

The K-nearest neighbor algorithm with cosine similarity search is utilized in MongoDB to retrieve the embedded data and subsequently transmit it to the LLM. The utilization of the K-nearest neighbor algorithm, which operates by identifying the shortest distance between the query and the instance data to produce training results, in conjunction with the cosine similarity search technique enhances the accuracy and efficiency of question production \cite{b17}.

The project's objective is to assist inexperienced coders in overcoming their coding phobia. The MVP incorporates an AI chatbot that aids users in the reduction of anxiety and the promotion of mental health by employing mindfulness techniques. Figure 5 depicts a sample query ``How do I know if I have depression," and the output generated by LLM using the trained dataset. Likewise, individuals have the ability to employ the AI chatbot functionality within the system to seek aid for any challenges they may face during the process of learning through the application. 

\begin{figure}[htbp]
\centerline{\includegraphics[width=8cm, height= 3cm]{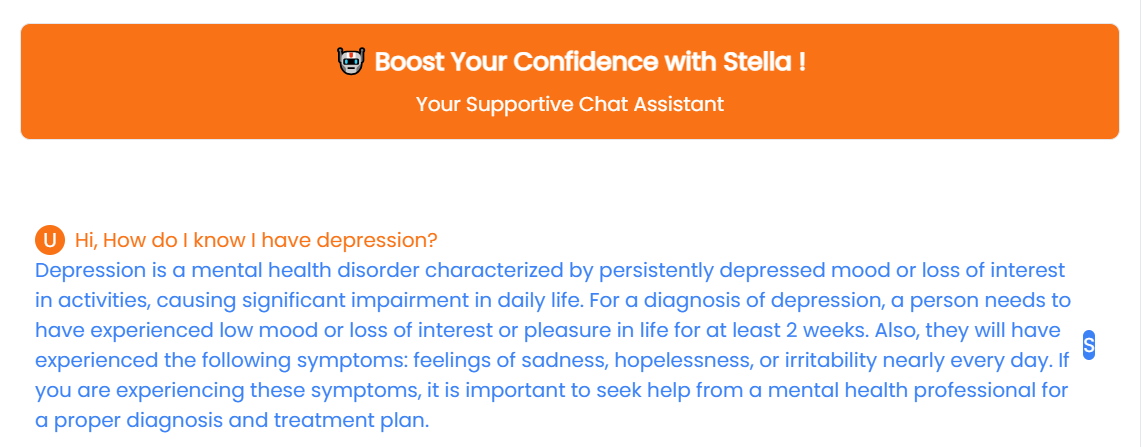}}
\caption{LLM generated chatbot response}
\label{figure}
\end{figure}

Subsequently, Figure 6 shows an additional exemplar query, ``I feel worthless, I can't code." While the LLM produces a moderately accurate response, it fails to deliver the most anticipated accurate answer due to the absence of a dataset specifically designed to address coding anxiety issues, as previously stated. In subsequent advancements, the author intends to integrate a distinct dataset on coding anxiousness in order to obtain more precise generations from the LLM.

\begin{figure}[htbp]
\centerline{\includegraphics[width=8cm, height=1.5cm]{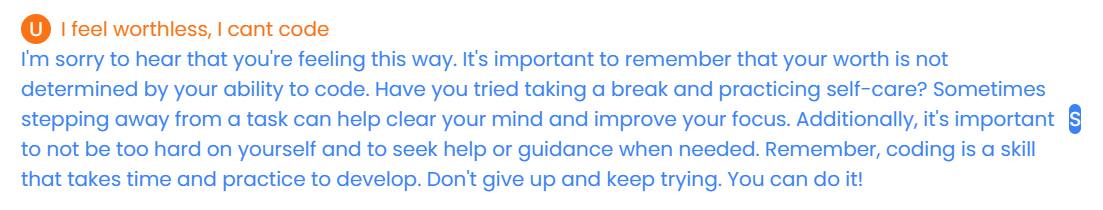}}
\caption{LLM generated chatbot response}
\label{figure}
\end{figure}

The ``Tip of the Day" is an additional feature that promotes mindfulness approach in the solution. Generative AI is employed by the LLM to generate motivational suggestions that serve to enhance the user's drive during system usage. This feature updates on a daily basis and is helpful for users to maintain confidence throughout their learning journey. Figure 7 showcases a sample output retrieved from the LLM, which includes the motivational tip: ``Practice deep breathing exercises to help stay calm and focused while coding."

\begin{figure}[htbp]
\centerline{\includegraphics[width=5cm, height=3cm]{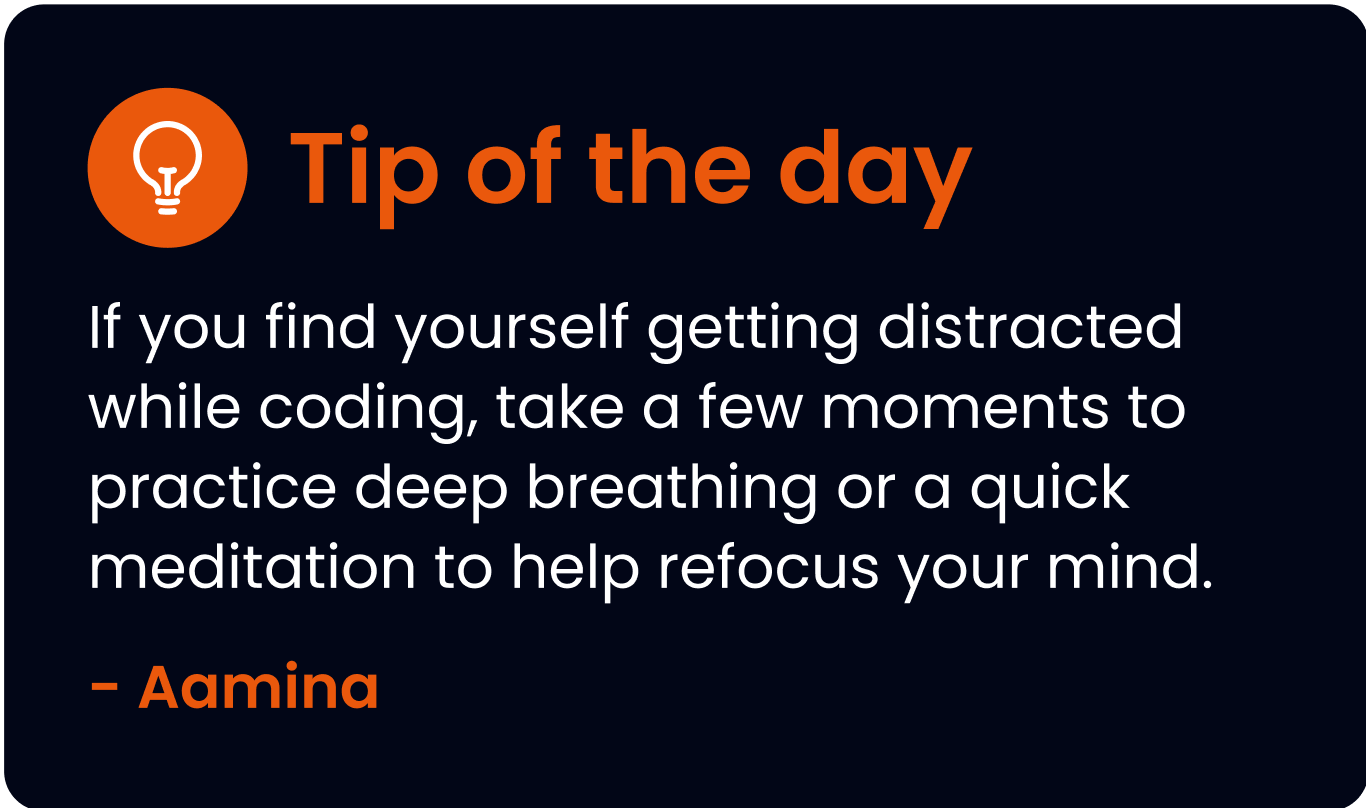}}
\caption{``Tip of the Day" feature}
\label{figure}
\end{figure}

The implementations indicated above, which utilize OpenAI, LangChain, and MongoDB, provide a significant illustration of the functional prototype and showcase the author's incorporation of generative AI into educational processes through the use of RAG-based LLM. It is expected that the implementation will be further refined, evaluated, and the initiative will be developed further. 

\section{Conclusion}
In summary, this study integrates the implementation of a code learning platform aimed at assisting novice coders in overcoming their fear of coding through the integration of mindfulness approaches and gamification elements using Generative AI. The minimum viable product has been developed using the RAG-based OpenAI LLM as the Generative AI model, and the development process and evaluation are continually being improved to further enhance the learning experience of users. Accodemy \cite{b22} represents a novel approach whereby novice coders can enjoy the learning process while prioritizing mental health, thus reducing anxiety associated with programming education.

\section*{Acknowledgment}
The authorship team would like to acknowledge the vision, support and guidance of the IEEE Industrial Electronics Society in conducting the Generative AI Hackathon under the leadership of Dr. Daswin De Silva and Lakshitha Gunasekara.


\begin{thebibliography}{00}
\bibitem{b1} A. Koyuncu and B. Koyuncu, “The Universal Skill of 21 st Century, Coding and Attitude of Secondary School Students towards Coding,” vol. 11, pp. 68–80, 2019. Accessed: Mar. 15, 2024. [Online].  Available at: https://files.eric.ed.gov/fulltext/EJ1325994.pdf 
\bibitem{b2} S. Santos, P. Tedesco, M. Borba, and M. Brito, “Innovative Approaches in Teaching Programming: A Systematic Literature Review,” Proceedings of the 12th International Conference on Computer Supported Education, 2020.  Accessed: Mar. 15, 2024. [Online].
Available at: https://doi.org/10.5220/0009190502050214
\bibitem{b3} J. Bennedsen and M. E. Caspersen, “Failure rates in introductory programming,” ACM SIGCSE Bulletin, vol. 39, no. 2, p. 32, Jun. 2007. Accessed: Mar. 15, 2024. [Online]. 
Available at: https://doi.org/10.1145/1272848.1272879
\bibitem{b4} B. Özmen and A. Altun, “Undergraduate Students’ Experiences in Programming: Difficulties and Obstacles,” Turkish Online Journal of Qualitative Inquiry, vol. 5, no. 3, Mar. 2014. Accessed: Mar. 15, 2024. [Online]. Available at: https://doi.org/10.17569/tojqi.20328
\bibitem{b5} C. Connolly, E. Murphy, and S. Moore, “Programming Anxiety Amongst Computing Students—A Key in the Retention Debate?,” IEEE Transactions on Education, vol. 52, no. 1, pp. 52–56, Feb. 2009. Accessed: Mar. 15, 2024. [Online].  Available at: https://doi.org/10.1109/TE.2008.917193 
\bibitem{b6} M. J. Scott, “Self-beliefs in the introductory programming lab and game-based fantasy role-play,” bura.brunel.ac.uk, 2015. Accessed: Mar. 19, 2024. [Online].  Available at: http://bura.brunel.ac.uk/handle/2438/11047
\bibitem{b7} S. E. Chang, “Computer anxiety and perception of task complexity in learning programming-related skills,” Computers in Human Behavior, vol. 21, no. 5, pp. 713–728, Sep. 2005. Accessed: Mar. 19, 2024. [Online].  Available at: https://doi.org/10.1016/j.chb.2004.02.021
\bibitem{b8} V. J. Shute and F. Ke, “Games, Learning, and Assessment,” Assessment in Game-Based Learning, pp. 43–58, 2012. Accessed: Mar. 15, 2024. [Online]. Available at: https://doi.org/10.1007/978-1-4614-3546-4\_4
\bibitem{b9} Jiang, Yongling, Z. Zhao, L. Wang, and S. Hu. "Research on the Influence of Technology-Enhanced Interactive Strategies on Programming Learning." Presented at the 2020 15th International Conference on Computer Science \& Education (ICCSE), IEEE, 2020. Accessed: Mar. 15, 2024. [Online].  Available at: https://ieeexplore.ieee.org/document/9201627
\bibitem{b10} B. Poellhuber, M. Chomienne, and T. Karsenti, “The Effect of Peer Collaboration and Collaborative Learning on Self-Efficacy and Persistence in a Learner-Paced Continuous Intake Model,” vol. 22, no. 3, pp. 41–62, 2008. Accessed: Mar. 15, 2024. [Online]. Available at: https://files.eric.ed.gov/fulltext/EJ812561.pdf 
\bibitem{b11} Khan Academy, “Khan academy,” Khan Academy, 2023. Accessed: Apr. 05, 2024. [Online]. Available at: https://www.khanacademy.org/
\bibitem{b12} S. Mondal, A. Gugnani, R. Sindhgatta, and V. K. R. Kasireddy, “Khan Academy: A Social Networking and Community Question Answering Perspective,” presented at the 2018 IEEE International Conference on Data Mining Workshops (ICDMW), IEEE, Feb. 2019. Accessed: Mar. 05, 2024. [Online]. Available at: https://ieeexplore.ieee.org/document/8637531
\bibitem{13} “Learn to Code - for Free,” Codecademy. Accessed: Apr. 05, 2024. [Online]. Available at: https://www.codecademy.com/
\bibitem{b14} R. Shen and M. J. Lee, “Learners’ Perspectives on Learning Programming from Interactive Computer Tutors in a MOOC,” presented at the 2020 IEEE Symposium on Visual Languages and Human-Centric Computing (VL/HCC), IEEE, Jul. 2020. Accessed: Mar. 05, 2024. [Online]. Available at: https://ieeexplore.ieee.org/document/9127270
\bibitem{b15} C. Jeong, “A Study on the Implementation of Generative AI Services Using an Enterprise Data-Based LLM Application Architecture,” arXiv.org, Sep. 18, 2023. Accessed: Mar. 15, 2024. [Online]. Available at: https://arxiv.org/abs/2309.01105
\bibitem{b16} H. Steck, C. Ekanadham, and N. Kallus, “Is Cosine-Similarity of Embeddings Really About Similarity?,” arXiv (Cornell University), Mar. 2024. Accessed: Mar. 19, 2024. [Online].  Available at: https://doi.org/10.1145/3589335.3651526
\bibitem{b17} M. Nursalman, J. Kusnendar, and U. F. Fadhila, “Implementation of K-Nearest Neighbor with Cosine Similarity for Classification Abstract International Journal of Computer Science,” presented at the 2018 International Conference on Information Technology Systems and Innovation (ICITSI), IEEE Oct. 2018. Accessed: Mar. 15, 2024. [Online]. Available at: https://doi.org/10.1109/icitsi.2018.8696072 
\bibitem{b18} O. A. Resta, A. Aditya, and F. E. Purwiantono, “Plagiarism Detection in Students’ Theses Using The Cosine Similarity Method,” SinkrOn, vol. 5, no. 2, pp. 305–313, May 2021. Accessed: Mar. 19, 2024, [Online]. Available at: doi: https://doi.org/10.33395/sinkron.v5i2.10909.
\bibitem{b19} K. Guo, “Testing and Validating the Cosine Similarity Measure for Textual Analysis,” SSRN Electronic Journal, 2022. Accessed: Mar. 15, 2024. [Online]. Available at: https://doi.org/10.2139/ssrn.4258463 
\bibitem{b20} J. Dumalo, “Chatbot for Mental Health Conversations,” kaggle.com. Accessed: Mar. 5, 2024. [Online]. Available at: www.kaggle.com/code/jocelyndumlao/chatbot-for-mental-health-conversations
\bibitem{b21} S. Barnett, S. Kurniawan, Srikanth Thudumu, Z. Brannelly, and M. Abdelrazek, “Seven Failure Points When Engineering a Retrieval Augmented Generation System,” arXiv.org (Cornell University), Jan. 2024. Accessed: Mar. 15, 2024. [Online]. Available at: https://doi.org/10.48550/arxiv.2401.05856
\bibitem{b22} accodemy.tech (2024). Accodemy web application Accessed: Mar. 23, 2024. [online] Available at: https://www.accodemy.tech/home 
\end{thebibliography}
\end{document}